# Research Title

Diagnosing Bipolar Disorder from 3-D Structural Magnetic Resonance Images Using a Hybrid GAN-CNN Method

# Authors and Affiliations

Masood Hamed Saghayan[1], Mohammad Hossein Zolfagharnasab[1], Ali Khadem[2,*], Farzam Matinfar[3], Hassan Rashidi[4]

[1]Master of Science, Faculty of Statistics, Mathematics, and Computer Science, Allameh Tabataba'i, Tehran, Iran
[2]Assistant Professor, Department of Biomedical Engineering, Faculty of Electrical Engineering, K. N. Toosi University of Technology, Tehran, Iran
[3]Assistant Professor, Faculty of Statistics, Mathematics, and Computer Science, Allameh Tabataba'i, Tehran, Iran
[4]Professor, Faculty of Statistics, Mathematics, and Computer Science, Allameh Tabataba'i, Tehran, Iran
[*]Corresponding Author: alikhadem@kntu.ac.ir

# Statements and Declarations

The authors of this paper declare that they have no known competing financial interests or personal relationships that could have appeared to influence the work reported in this paper. Also, the authors did not receive support from any organization for the submitted work.

# Acknowledgments

The authors carry out the current research, and no organization or funding supports this research. The contributors are Masood Hamed Saghayan, Mohammad Hossein Zolfagharnasab, Ali Khadem, Farzam Matinfar, and Hassan Rashidi.

# Abstract

Bipolar Disorder (BD) is a psychiatric condition diagnosed by repetitive cycles of hypomania and depression. Since diagnosing BD relies on subjective behavioral assessments over a long period, a solid diagnosis based on objective criteria is not straightforward.

The current study responded to the described obstacle by proposing a hybrid GAN-CNN model to diagnose BD from 3-D structural MRI Images (sMRI). The novelty of this study stems from diagnosing BD from sMRI samples rather than conventional datasets such as functional MRI (fMRI), electroencephalography (EEG), and behavioral symptoms while removing the data insufficiency usually encountered when dealing with sMRI samples. The impact of various augmentation ratios is also tested using 5-fold cross-validation.

Based on the results, this study obtains an accuracy rate of 75.8%, a sensitivity of 60.3%, and a specificity of 82.5%, which are 3-5% higher than prior work while utilizing less than 6% sample counts. Next, it is demonstrated that a 2-D layer-based GAN generator can effectively reproduce complex 3D brain samples, a more straightforward technique than manual image processing. Lastly, the optimum augmentation threshold for the current study using 172 sMRI samples is 50%, showing the applicability of the described method for larger sMRI datasets. In conclusion, it is established that data augmentation using GAN improves the accuracy of the CNN classifier using sMRI samples, thus developing more reliable decision support systems to assist practitioners in identifying BD patients more reliably and in a shorter period.

# Keywords

# Diagnosing Bipolar Disorder from 3-D Structural MagneticResonance Images Using a Hybrid GAN-CNN Model

Masood Hamed Saghayan[1], Mohammad Hossein Zolfagharnasab[1],

Ali Khadem[2,*], Farzam Matinfar[1], Hassan Rashidi[1]

[1]*Faculty of Statistics, Mathematics, and Computer Science, Allameh Tabataba'i, Tehran, Iran.*

[2]*Department of Biomedical Engineering, Faculty of Electrical Engineering, K. N. Toosi University of Technology, Tehran, Iran.*

[*]*Corresponding Author*: alikhadem@kntu.ac.ir

## Abstract

Bipolar Disorder (BD) is a psychiatric condition diagnosed by repetitive cycles of hypomania and depression. Since diagnosing BD relies on subjective behavioral assessments over a long period, a solid diagnosis based on objective criteria is not straightforward. The current study responded to the described obstacle by proposing a hybrid GAN-CNN model to diagnose BD from 3-D structural MRI Images (sMRI). The novelty of this study stems from diagnosing BD from sMRI samples rather than conventional datasets such as functional MRI (fMRI), electroencephalography (EEG), and behavioral symptoms while removing the data insufficiency usually encountered when dealing with sMRI samples. The impact of various augmentation ratios is also tested using 5-fold cross-validation.

Based on the results, this study obtains an accuracy rate of 75.8%, a sensitivity of 60.3%, and a specificity of 82.5%, which are 3-5% higher than prior work while utilizing less than 6% sample counts. Next, it is demonstrated that a 2-D layer-based GAN generator can effectively reproduce complex 3D brain samples, a more straightforward technique than manual image processing. Lastly, the optimum augmentation threshold for the current study using 172 sMRI samples is 50%, showing the applicability of the described method for larger sMRI datasets. In conclusion, it is established that data augmentation using GAN improves the accuracy of the CNN classifier using sMRI samples, thus developing more reliable decision support systems to assist practitioners in identifying BD patients more reliably and in a shorter period.

## 1 Introduction

Recent developments in artificial intelligence (AI) technology have proven effective in improving decision support systems in psychiatric applications. (Zhang et al., 2021). As there are few behavior-independent tools for detecting mental illness, modern AI models such as Deep Neural Networks (DNN) can assist practitioners in making informed decisions in a shorter period (Lu et al., 2021). However, developing such systems requires resolving several main obstacles.

Firstly, modern AI models require significantly more training samples to obtain maturity than their less complicated predecessors, such as Supported Vector Machine (SVM) (Browarczyk et al., 2020). For instance, even simple neural network models used in imaging applications nowadays contain thousands to millions of parameters that must be accurately tuned during the training session (Saghayan et al., 2021). Therefore, the data insufficiency in modern AI models is more intense, especially in medical applications in which large datasets are not generally available due to the high cost of the operations, patient confidentiality, and the inability to repeat a given test in a short period (Schliebs & Kasabov, 2013). Consequently, collecting sufficient samples in medical applications can pose a significant challenge when developing fine-tuned decision support systems (Yu et al., 2021).

Secondly, the training process in modern AI models is not straightforward due to their growing complexity and hybrid functionality in recent years (Tanveer et al., 2022). For instance, models based on Convolutional Neural Networks (CNN) consist of initial feature extraction layers, significantly reducing the manual preprocessing task used in primitive Machine Learning (ML) methods to detect impactful parameters. On the downside, increasing the model complexity makes the training process time-consuming (Ge et al., 2022). As a result, developing an AI model with advanced functionality is challenging, especially for medical purposes in which reliability is essential. To conclude the discussions mentioned above, the two major issues concerning the application of modern AI models in medical applications can be summarized as having very few behavioral-independent tools to detect mental disorders and insufficient sample counts to train complex AI models. The current study responds to the described problems by the following objectives.

Initially, this study evaluates using sMRI samples to assess Bipolar Disorder (BD) from behavioral-independent medical samples. Compared with the conventional data formats such as fMRI, EEG, and other physical examinations, while sMRI datasets are categorized as a high-resolution biomarker, they are less prevalent in building Deep Learning (DL) due to the requiring more sample counts to obtain a well-developed model (Emmert-Streib et al., 2020). The current study removes this obstacle by proposing a hybrid pipeline in which the sMRI samples are augmented using a Generative Adversarial Network (GAN), thus improving the classifier predictions. It is also worth noting that developing such frameworks not only removes the data insufficiency explained earlier but also replicates valid medical samples, which conventional image processing methods are most likely to fail as the sample complexity grows.

Next, the computational load of training the whole three-dimensional model of sMRI images can rapidly cross the intolerable margin of many computational units. The current study resolves this issue by constructing the brain samples as individual two-dimensional layers. These sublayers produced by the GAN generator are then assembled into the high-resolution full-scale sMRI data representing the brain structure. Evaluating such attributes is most demanding since working through two-dimensional sub-models reduces the computational load to a tolerable threshold without performing destructive approaches such as direct down-sampling, which is the commonly used technique when dealing with extensive data (Chalehchaleh & Khadem, 2021).

The remaining of this study is divided into seven sections. In *Section 2*, a summary of the related works is presented. *Section 3* provides the necessary information regarding the selected dataset. Next, *Section 4* explains the solution pipeline comprising preprocessing procedure, GAN architecture, and the classifier properties. The classifier metrics, comparison with prior studies, and the results related to the sample generation are presented in *Section 5*. The discussions regarding the impact of data augmentation on classification performance are carried out in *Section 6*. The limitations of this study are described in *Section 7*, followed by recommendations for future work in *Section 8*. Finally, the conclusions are summarized in *Section 9*.

## 2 Related Works

Since BD patients tend to participate in destructive activities, recognizing BD from normal depression is crucial (Gojkovich & Rivardo, 2021). As a result, a diverse range of biomarkers, such as blood chemical samples, behavioral records, voice-tone analyses, electroencephalograph signals, and brain MRI samples, has been selected to detect BD symptoms (Goerigk et al., 2021).

As one of the pioneering works in the field, Nunes et al. (Nunes et al., 2020) developed an SVM model to diagnose BD for MRI images. Their investigation included MRI shots of 853 BD patients and 2167 normal individuals. Due to the inability of ML techniques for feature extraction, features such as cortical thickness, subcortical structures, the inferior gyrus, frontal gyrus, hippocampus, and amygdala were manually selected to train the SVM model. Based on their reports, their SVM model has obtained an accuracy of 65.23%, a sensitivity of 66.02%, and a specificity of 64.90%. Similarly, studies such as (Lee et al., 2020) and (Wu et al., 2022) have also developed SVM models capable of distinguishing BD patience with higher accuracy values. However, manual feature selection is found to be their mutual liability. Due to such drawbacks, Martyn et al. (Martyn et al., 2019) utilized the VGG16 model, a robust and well-tested CNN architecture that utilizes convolutional layers for automated feature extraction. As a result, they achieved an accuracy of 65% with sample counts less than 10% of prior SVM models. Unfortunately, a solid conclusion cannot be made on their developed model due to the absence of other evaluation parameters, such as sensitivity and specificity. Their dataset consists of three MRI sample sets with an overall count of 156 healthy individuals and 91 patients suffering from BD. More information can be found in (Martyn et al., 2019).

Lastly, many other studies such as (Lei et al., 2022; Mateo-Sotos et al., 2022; Weissmann et al., 2020) have also detected BD patients from healthy individuals with much higher accuracy bounds; however, their models are trained based on biomarkers other than MRI images. As a result, they are not relevant for further exploration.

## 3 Dataset

The dataset used in this study is a shared neuroimaging dataset obtained by the UCLA Consortium for Neuropsychiatric Phenomics (Poldrack et al., 2016). The dataset comprised a high angular resolution diffusion sMRI formatted according to the Brain Imaging Data Structure (BIDS) standard. Fig. (1) depicts a sample of sMRI images shown from multiple viewpoints, including the number of layers available from each view. More information regarding the dataset is presented in Table (1).

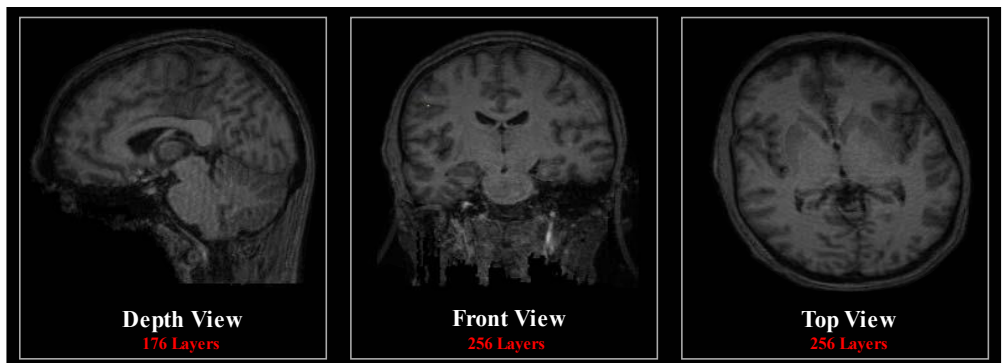

**Fig. (1) Schematic functionality of GAN models.**

Table 1- Dataset Information.

| Dataset Properties | Bipolar Disorder | Normal |
| --- | --- | --- |
| Number of Samples | 49 | 123 |
| Train Samples | 37 | 92 |
| Test Samples | 12 | 31 |
| Age-Range | 21-50 | 22-50 |
| Sex Ratio (Men/Women) | 1.33 | 1.09 |
| Type of Imaging | Structured MRI T1-Weighted (*.nii) | Structured MRI T1-Weighted (*.nii) |
| Data Dimension | (Front×Top×Depth) = 256×256×172 | (Front×Top×Depth) = 256×256×172 |

# 4 Method

The workflow followed in this study is divided into several steps: data preparation (preprocess), sample augmentation, classifier training, and model evaluation. For the convenience of future readers, a schematic flowchart is also provided in Fig. (2). More essential points regarding each step are described in the following sections.

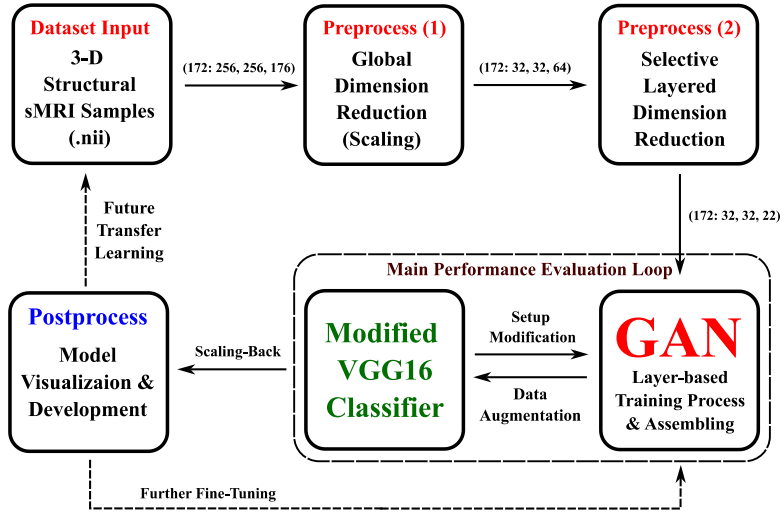

Fig. (2) Study workflow.

## 4.1 Preprocessing

The preprocessing stage is divided into two separate steps. Initially, a uniform dimension reduction shrunk the original data size by *128* times from *(256×256×176)* to *(64×64×44)* such that the computational load is reduced to a tolerable measure. This process is called global dimension reduction, in which all sample layers are affected equally. Next, a non-uniform layer-based dimension reduction is carried out such that only the most informative layers (22 middle layers) are kept for the training session, and layers without perceptible brain structure are discarded to reduce the computational load. This issue can be vividly observed by comparing the brain structure depicted in Fig (3) at each depth.

As a result, it can be concluded that discarding the farther layers is a reasonable down-scaling while minimizing the loss of impactful information. It must be noted that each of the preprocessing steps can be adapted for computational units with higher capacity.

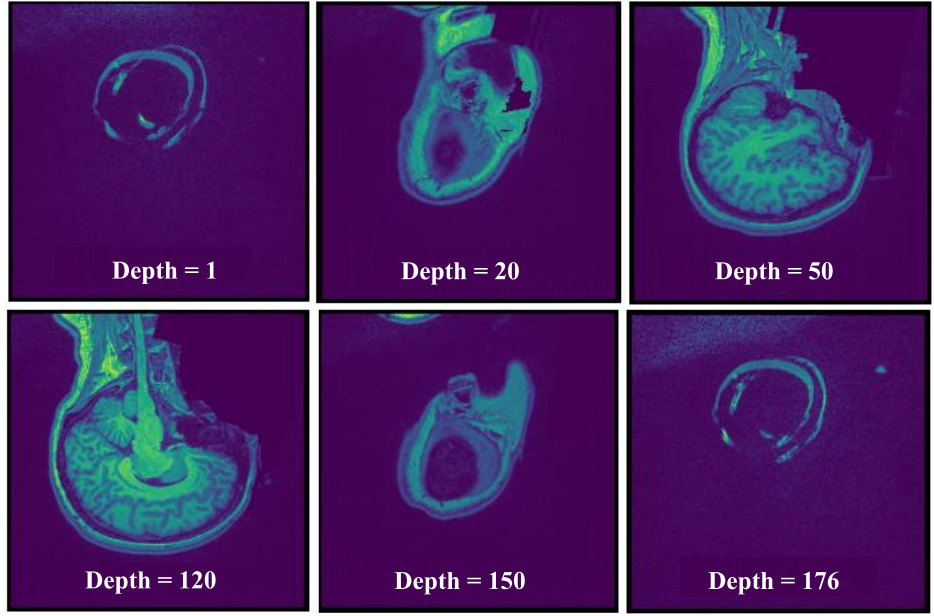

**Fig. (3) Brain structure visibility in sMRI depth layers.**

## 4.2 Data Augmentation

A compact schematic of GAN functionality is depicted in Fig. (4). Technically, GAN is defined as an indirect training of a generator and a discriminator in a zero-sum game in which a generator produces fake samples from a normally distributed noise, while the discriminator distinguishes the fake outputs from the original samples (You et al., 2022). Although the discriminator requires fewer training epochs than the generator, an optimum design directly affects the quality of sample generation (Ahmad et al., 2022). More information regarding the GAN properties is provided in Table (2).

Since the computational load required to train all layers is not tolerable, an individual GAN model is trained to generate each depth layer separately. Next, these generated layers are assembled into a full-scaled model representing the entire brain structure. The noted strategy is also beneficial in reducing the sample counts required to achieve a well-trained network.

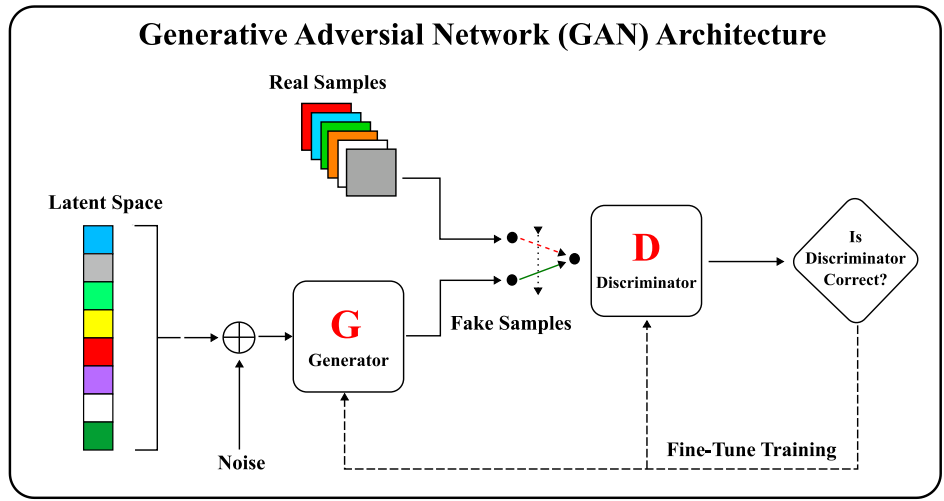

**Fig. (4) Schematic structure of GAN models.**

Table 2- Configuration of GAN Layers.

| Type | Number | Layer Properties | Output Dimension | Parameter Count |
|---|---|---|---|---|
| Generator | 1 | Dense | 1024 | 512000 |
| | 2 | Batch Normalization | 1024 | 4096 |
| | 3 | Leaky-ReLU | 1024 | 0 |
| | 4 | Dense | 65536 | 1.09 |
| | 5 | Leaky-ReLU | 65536 | 0 |
| | 6 | Reshape | 16×16×256 | 0 |
| | 7 | Convolutional Layer (2D) | 32×32×64 | 409600 |
| | 8 | Batch Normalization | 32×32×64 | 256 |
| | 9 | Leaky-ReLU | 32×32×64 | 0 |
| | 10 | Convolutional Layer (2D) | 64×64×1 | 1600 |
| | **Total Number of Parameters: 68,101,952** | | | |
| Discriminator | 1 | Convolutional Layer (2D) | 32×32×64 | 1664 |
| | 2 | Leaky-ReLU | 32×32×64 | 0 |
| | 3 | Dropout | 32×32×64 | 0 |
| | 4 | Convolutional Layer (2D) | 16×16×128 | 204928 |
| | 5 | Leaky-ReLU | 16×16×128 | 0 |
| | 6 | Dropout | 16×16×128 | 0 |
| | 7 | Flatten | 32768 | 0 |
| | 8 | Dense | 64 | 2097216 |
| | 9 | Dense | 64 | 4160 |
| | 10 | Dense | 1 | 65 |
| | **Total Number of Parameters: 2,308,033** | | | |

## 4.3 Classifier Properties

This study selects a binary-type CNN model to detect BD patients from healthy individuals. The submodel used for this purpose is the famous VGG16 with minor tuning to improve the classification performance. The network inputs are grayscale sMRI images with $32 \times 32 \times 22$ dimensions containing real and fake train samples. Also, a 5-fold cross-validation is performed to evaluate the classification performance. More information regarding the classifier architecture can be found in Table (3).

Table 3- Configuration of Classifier Layer Properties.

| Number | Layer Properties | Output Dimension | Parameter Count |
|---|---|---|---|
| 1 | Convolutional Layer (3D) | 30×30×20×64 | 1792 |
| 2 | Activation | 30×30×20×64 | 0 |
| 3 | Max Pooling (3D) | 15×15×10×64 | 0 |
| 4 | Batch Normalization | 15×15×10×64 | 256 |
| 5 | Convolutional Layer (3D) | 13×13×8×64 | 110656 |
| 6 | Activation | 13×13×8×64 | 0 |
| 7 | Max Pooling (3D) | 6×6×4×64 | 0 |
| 8 | Batch Normalization | 6×6×4×64 | 256 |
| 9 | Convolutional Layer (3D) | 4×4×2×64 | 110656 |
| 10 | Activation | 4×4×2×64 | 0 |
| 11 | Flatten | 2048 | 0 |
| 12 | Dense | 1024 | 2098176 |
| 13 | Activation | 1024 | 0 |
| 14 | Dense | 256 | 262400 |
| 15 | Activation | 256 | 0 |
| 16 | Dense | 2 | 514 |
| **Total Number of Parameters: 2,584,706** | | | |

# 5. Results

In general, sMRI samples are an ideal biomarker to diagnose BD patients only if the brain structure is affected by the BD. However, since BD does not necessarily leave a mark on the brain, classifying patients using sMRI samples requires extra caution.

In the current study, no extra label is considered in the dataset to specify whether the BD has changed the brain structure; therefore, it is assumed that BD affected all the corresponding samples. As anticipated, the noted assumption is inaccurate since the value success rate of detecting BD patients (the sensitivity) is 60.3%, much lower than the success rate of predicting normal individuals (specificity rate), which is about 82.5%. Nonetheless, the performance of the developed model is found reasonable for samples in which the BD has reflected the slightest mark on the brain. Therefore, an accuracy rate of 75.8% is much higher than the obtained sensitivity values.

## 5.1 Comparison with Prior Works

This study is compared with (Nunes et al., 2020), in which a similar procedure consisting of data augmentation and sample classification is carried out to detect BD from normal individuals. One of the critical differences with (Nunes et al., 2020) is that the current study selects models based on neural networks to perform both augmentation and classification tasks, whereas (Nunes et al., 2020) select statistical approaches to perform similar tasks. The noted decision increases the potential of the current model to achieve more maturity. For instance, the sensitivity and specificity rates of (Nunes et al., 2020) are very close. Considering that their dataset is 20 times bigger than the one selected in this study, it can be established that their model achieved its highest limit in classifying normal and BD individuals.

On the other hand, the specificity of the current model is comparably higher than the sensitivity value, which demonstrates the high capability of the presented model in detecting BD for cases in which the illness has left the slightest impact on the brain structure. Therefore, it can be concluded that the presented framework is sufficiently accurate for further development as a decision-support system that can assist practitioners in daily clinical usage.

**Table 4- Comparison to prior study (Nunes et al., 2020).**

| Model | Summary | Increasing | Classifier | Sensitivity | Specificity | Precision | Accuracy | F1-score |
|---|---|---|---|---|---|---|---|---|
| Current study | 123 Normal 49 Bipolar | GAN | CNN | 60.3% | 82.5% | 55.2% | 75.8% | 57.1% |
| Ref. | 853 Normal 2167 Bipolar | Aggregated | SVM | 63.2% | 64.9% | 44.4% | 65.2% | 52.3% |

## 5.2 Data Augmentation

As discussed earlier, the current study resolves the sample deficiency using the GAN model. Fig. (5) presents the maturity of the GAN model throughout the training. The maximum number of training epochs is selected as 20,000 to guarantee that the model is well-trained; however, it is observed that the GAN outputs do not vary after the initial 8,000 training iterations.

Based on the results, the GAN model produced nothing but pure noises during the first few iterations, demonstrating that the weights corresponding to the GAN generator are tuned from scratch. Afterward, a silhouette structure of the brain is observed in epoch 50, and the subsequent thousand epochs are spent fine-tuning the generator to produce a more realistic brain structure. As shown in Fig. (5), the brain image generated by GAN shows only the blear appearance in epoch 1000; however, the image corresponding to iteration 10,000 illustrates a vivid brain structure.

As a final note, It must be stated that although the GAN generator is the only part used further in this study to replicate fake samples, the discriminator unit is the part that required more epochs to achieve a well-trained status. Therefore, more attention should be reflected to remove convergence difficulties associated with the discriminator.

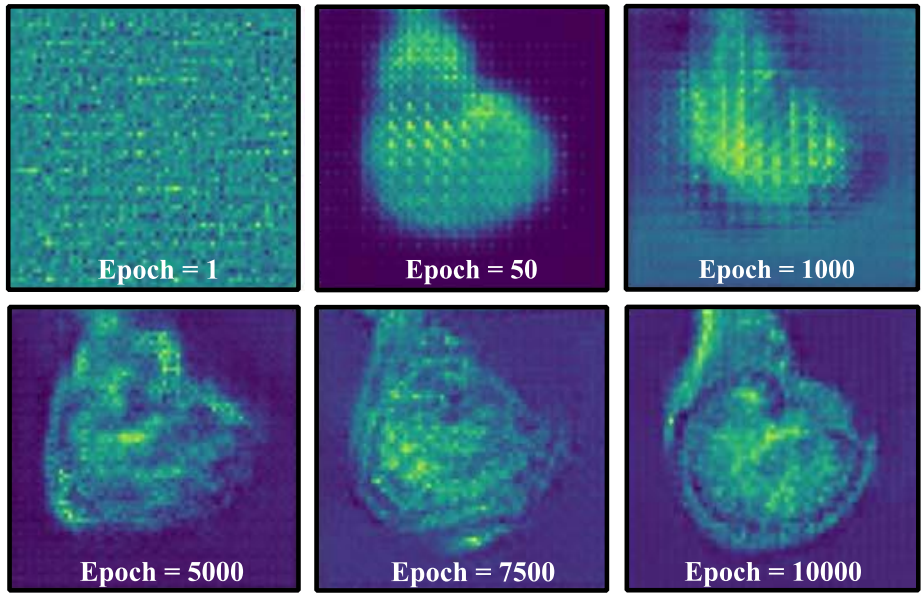

**Fig. (5) GAN predictions of brain structure during the generator training.**

## 6. Discussion

Now that the accuracy of the current model is established, the subsequent issue of interest is to investigate the impact of data augmentation on the classifier's performance. However, three critical issues are worth discussing before going through the investigation. Firstly, a positive impact of sample augmentation confirms the effectiveness of the layered base (bottom-up) GAN training used in this study. Secondly, obtaining an optimum threshold for data augmentation is crucial since over-using augmented samples not only reflects a significant computational load on the solution procedure but also reduces the overall accuracy of the model predictions. Lastly, the augmented samples are not used for testing since the model might be biased as the number of fake samples grows. Doing so increases the reliability of the test phase since the number of test samples is consistent among all data augmentation scenarios, and the classifier would not be biased as the number of fake samples grows.

Accordingly, various augmentation ratios by up to 300% data increase are selected, and the classifier metrics are recorded in Table (5). The results show that the augmentation ratio positively impacts the classification performance. For instance, an augmentation ratio of 25% improves the accuracy rate by 4% and the F1-score by almost 8% compared with the base scenario (no sample augmentation).

Next, increasing the augmentation ratio to 50% increases the accuracy rate by 6%, and the F1 score is increased by 13% compared with the base scenario, demonstrating the GAN network's potential in replicating brain samples. Unfortunately, further increasing the augmentation ratio decreases accuracy metrics. For example, by excessively increasing the sample augmentation up to 300% compared with the base scenario, the F1-score is decreased by 2% while the accuracy rate remains almost the same as the no augmentation case. This observance proves that the model prediction deteriorates as the number of generated samples increases after a certain threshold.

As a result, it can be concluded that the necessity to provide sufficient sample counts can not be entirely omitted through GAN augmentation. In this regard, even though utilizing GAN is a practical solution to improve the performance of AI classifiers and compensate for data insufficiency, an individual case study is still required to estimate the exact criterion. In the current investigation, the noted threshold is 50% of the sample counts, which can be considered a safe margin for a wide range of medical imaging applications, considering the complexity of constructing brain samples. Furthermore, the success of data augmentation in the current investigation points to the validity of the layered-based training of the GAN, which has yielded the accurate resemblance of 3-D brain samples even when a discrete bottom-up approach is carried out to assemble 2-D models to construct a full-scale model. Lastly, the increase in F1-score indicates that the data augmentation increased the number of samples in which the BD has left a mark on the brain structure for the classifier to detect. While such brain marks can only be observed by an expert eye, the increase in F1-score is an indirect sign that the number corresponding to the noted samples is increasing due to them being responsible for improving the classifier's predictions.

Table 5- Impact of Data Augmentation on the Classifier's Overall Performance.

| Augmentation Ratio | Base-0% | 25% | 50% | 75% | 100% | 300% |
|---|---|---|---|---|---|---|
| Normal (Train, Test) | (92, 31) | (122, 31) | (153, 31) | (184, 31) | (215, 31) | (474, 31) |
| Bipolar (Train, Test) | (37, 12) | (49, 12) | (61, 12) | (73, 12) | (85, 12) | (181, 12) |
| Accuracy rate | 69.4% | 74.8% | 75.8% | 73.6% | 72.4% | 69.5% |
| F1-score | 44.2% | 52.3% | 57.1% | 51.8% | 48.6% | 42.4% |

## 7 Limitations

There are three main limitations involved in the current study. First, the selected dataset has no extra label indicating whether or not the BD has left a viable mark on the brain sMRI sample. Since the classifier can only detect BD when the brain structure is affected by the brain, BD patients without a visible mark on their brain are not expected to be classified correctly. Second, the available computational unit is found to be insufficient compared with the high dimensionality of the samples. The described obstacle is solved by applying a uniform dimension reduction and selecting brain layers with the highest rate of information to shrink the computational load to a tolerable threshold. Although operations such as down-scaling and selective training may compulsorily remove some portion of sample features, it facilitates all model development processes by reducing the neural network parameters (model complexity) to achieve a well-developed model. Lastly, the hyper-parameters involved in this study are not perfectly tuned. In this regard, future studies can fine-tune these variables to obtain more accurate results depending on their dataset properties and sample dimension.

## 8 Future Works

This study can be further developed from three different directions. First, this study can be extended to include similar mental disorders such as schizophrenia and major depression in diagnosis. Doing so would make the model more applicable in daily clinical usage. Second, the disadvantage of sMRI samples, which is the uncertainty that BD has left a mark on the brain, can be removed by coupling this model with other biomarkers through a voting mechanism responsible for selecting between the models based on their presented statistics. By doing so, the uncertainty regarding the sMRI samples is eliminated due to exposing more parameters to the model. In the meantime, benefits such as the high accuracy of sMRI samples in reflecting the slightest BD marks on the brain are also included. Finally, it is recommended that the accuracy of the current study would be further assessed on a potential BD subject with regular clinical follow-ups. By doing so, the predictions of the presented model can assist professionals in diagnosing BD patients with more certainty.

## 9. Concluding Remarks

Using CNN and GAN data augmentation on sMRI samples, the current study proposes a decision support system functioning independently of behavioral symptoms to detect Bipolar Disorder (BD) from brain images. Based on the obtained results, the conclusions are summarized as follows:

- ✓ It is found that deep neural network classifiers such as the one used in this study can accurately detect the mark that BD leaves on the brain structure.
- ✓ The modified VGG-16 classifier used in this study obtains an accuracy rate of 75.8%, a sensitivity of 60.3%, and a specificity of 82.5%, which are 3-5% higher than prior work compared with prior studies in which SVM is applied.
- ✓ The intolerable computational load required to train the 3-D sMRI samples is removed by training an individual GAN model for a selective number of depth layers. However, the results indicate that the assembled model accurately represents the entire brain structure, and the GAN model has matured after 10,000 epochs.
- ✓ The GAN model improves the classifier's performance by augmenting the available samples. Doing so shows an average 5% increase in classification metrics; however, a specific limit in using the augmented samples is found, which is 50% of the total in the current study. Therefore, GAN cannot fully compensate for the sample insufficiency.

In conclusion, it is found that with the least amount of sample counts, the combination of the GAN augmentation technique and CNN is capable of developing accurate decision support frameworks with the least amount of sample count. Therefore, the current model can be utilized as an additional behavioral-independent tool to guide practitioners to achieve a solid decision on BD diagnosis in a shorter period.

## Statements and Declarations

The authors of this paper declare that they have no known competing financial interests or personal relationships that could have appeared to influence the work reported in this paper. Also, the authors did not receive support from any organization for the submitted work.

## Acknowledgments

The authors carry out the current research, and no organization or funding supports this research. The contributors are Masood Hamed Saghayan, Mohammad Hossein Zolfagharnasab, Ali Khadem, Farzam Matinfar, and Hassan Rashidi.
## References

Ahmad, B., Sun, J., You, Q., Palade, V., & Mao, Z. (2022). Brain Tumor Classification Using a Combination of Variational Autoencoders and Generative Adversarial Networks. *Biomedicines*, *10*(2). https://doi.org/10.3390/biomedicines10020223

Browarczyk, J., Kurowski, A., & Kostek, B. (2020). Analyzing the effectiveness of the brain–computer interface for task discerning based on machine learning. *Sensors (Switzerland)*, *20*(8). https://doi.org/10.3390/s20082403

Chalehchaleh, A., & Khadem, A. (2021). Diagnosis of Bipolar i Disorder using 1 D-CNN and Resting-State fMRI Data. *Proceedings of the 5th International Conference on Pattern Recognition and Image Analysis, IPRIA 2021*. https://doi.org/10.1109/IPRIA53572.2021.9483574

Emmert-Streib, F., Yang, Z., Feng, H., Tripathi, S., & Dehmer, M. (2020). An Introductory Review of Deep Learning for Prediction Models With Big Data. In *Frontiers in Artificial Intelligence* (Vol. 3). https://doi.org/10.3389/frai.2020.00004

Ge, H., Zhu, Z., Dai, Y., Wang, B., & Wu, X. (2022). Facial expression recognition based on deep learning. *Computer Methods and Programs in Biomedicine*, *215*. https://doi.org/10.1016/j.cmpb.2022.106621

Goerigk, S., Cretaz, E., Sampaio-Junior, B., Vieira, É. L. M., Gattaz, W., Klein, I., Lafer, B., Teixeira, A. L., Carvalho, A. F., Lotufo, P. A., Benseñor, I. M., Bühner, M., Padberg, F., & Brunoni, A. R. (2021). Effects of tDCS on neuroplasticity and inflammatory biomarkers in bipolar depression: Results from a sham-controlled study. *Progress in Neuro-Psychopharmacology and Biological Psychiatry*, *105*. https://doi.org/10.1016/j.pnpbp.2020.110119

Gojkovich, K. L., & Rivardo, M. G. (2021). Effects of diagnosis and response style on social distance and perceived dangerousness. *North American Journal of Psychology*, *23*(3).

Hirschfeld, R. M. A., Lewis, L., & Vornik, L. A. (2003). Perceptions and impact of bipolar disorder: How far have we really come? Results of the National Depressive and Manic-Depressive Association 2000 Survey of individuals with bipolar disorder. In *Journal of Clinical Psychiatry* (Vol. 64, Issue 2). https://doi.org/10.4088/JCP.v64n0209

Lee, S. Y., Lu, R. B., Wang, L. J., Chang, C. H., Lu, T., Wang, T. Y., & Tsai, K. W. (2020). Serum miRNA as a possible biomarker in the diagnosis of bipolar II disorder. *Scientific Reports*, *10*(1). https://doi.org/10.1038/s41598-020-58195-0

Lei, Y., Belkacem, A. N., Wang, X., Sha, S., Wang, C., & Chen, C. (2022). A convolutional neural network-based diagnostic method using resting-state electroencephalograph signals for major depressive and bipolar disorders. *Biomedical Signal Processing and Control*, *72*. https://doi.org/10.1016/j.bspc.2021.103370

Lu, C. Y., Busch, A. B., Zhang, F., Madden, J. M., Callahan, M. X., LeCates, R. F., Wallace, J., Foxworth, P., Soumerai, S. B., Ross-Degnan, D., & Wharam, J. F. (2021). Impact of High-Deductible Health Plans on Medication Use Among Individuals With Bipolar Disorder. *Psychiatric Services (Washington, D.C.)*, *72*(8). https://doi.org/10.1176/appi.ps.202000362

Martyn, P., McPhilemy, G., Nabulsi, L., Martyn, F. M., Hallahan, B., McDonald, C., Cannon, D. M., & Schukat, M.

(2019). Using magnetic resonance imaging to distinguish a healthy brain from a bipolar brain: A transfer learning approach. *CEUR Workshop Proceedings*, *2563*.

Mateo-Sotos, J., Torres, A. M., Santos, J. L., Quevedo, O., & Basar, C. (2022). A Machine Learning-Based Method to Identify Bipolar Disorder Patients. *Circuits, Systems, and Signal Processing*, *41*(4). https://doi.org/10.1007/s00034-021-01889-1

Nunes, A., Schnack, H. G., Ching, C. R. K., Agartz, I., Akudjedu, T. N., Alda, M., Alnæs, D., Alonso-Lana, S., Bauer, J., Baune, B. T., Bøen, E., Bonnin, C. del M., Busatto, G. F., Canales-Rodríguez, E. J., Cannon, D. M., Caseras, X., Chaim-Avancini, T. M., Dannlowski, U., Díaz-Zuluaga, A. M., … Hajek, T. (2020). Using structural MRI to identify bipolar disorders – 13 site machine learning study in 3020 individuals from the ENIGMA Bipolar Disorders Working Group. *Molecular Psychiatry*, *25*(9). https://doi.org/10.1038/s41380-018-0228-9

Poldrack, R. A., Congdon, E., Triplett, W., Gorgolewski, K. J., Karlsgodt, K. H., Mumford, J. A., Sabb, F. W., Freimer, N. B., London, E. D., Cannon, T. D., & Bilder, R. M. (2016). A phenome-wide examination of neural and cognitive function. *Scientific Data*, *3*. https://doi.org/10.1038/sdata.2016.110

Saghayan, M. H., Seifpour, S., & Khadem, A. (2021). Automated Sleep Stage Scoring Using Brain Effective Connectivity and EEG Signals. *Proceedings - 2021 7th International Conference on Signal Processing and Intelligent Systems, ICSPIS 2021*. https://doi.org/10.1109/ICSPIS54653.2021.9729377

Schliebs, S., & Kasabov, N. (2013). Evolving spiking neural network-a survey. *Evolving Systems*, *4*(2). https://doi.org/10.1007/s12530-013-9074-9

Tanveer, M., Rajani, T., Rastogi, R., Shao, Y. H., & Ganaie, M. A. (2022). Comprehensive review on twin support vector machines. *Annals of Operations Research*. https://doi.org/10.1007/s10479-022-04575-w

Weissmann, D., Salvetat, N., Chimienti, F., Dupre, P., Dubuc, B., Cayzac, C., Checa Robles, F., & Patel, V. (2020). EDIT-B: A Blood Test to Diagnose Bipolar Disorder Using Epigenetic Biomarkers. *Biological Psychiatry*, *87*(9). https://doi.org/10.1016/j.biopsych.2020.02.716

Wu, X., Zhu, L., Zhao, Z., Xu, B., Yang, J., Long, J., & Su, L. (2022). Application of machine learning in diagnostic value of mRNAs for bipolar disorder. *Nordic Journal of Psychiatry*, *76*(2). https://doi.org/10.1080/08039488.2021.1937311

You, A., Kim, J. K., Ryu, I. H., & Yoo, T. K. (2022). Application of generative adversarial networks (GAN) for ophthalmology image domains: a survey. In *Eye and Vision* (Vol. 9, Issue 1). https://doi.org/10.1186/s40662-022-00277-3

Yu, Z., Amin, S. U., Alhussein, M., & Lv, Z. (2021). Research on Disease Prediction Based on Improved DeepFM and IoMT. *IEEE Access*, *9*. https://doi.org/10.1109/ACCESS.2021.3062687

Zhang, Y., Wang, S., Hermann, A., Joly, R., & Pathak, J. (2021). Development and validation of a machine learning algorithm for predicting the risk of postpartum depression among pregnant women. *Journal of Affective Disorders*, *279*.